\documentclass{emulateapj}

\shorttitle{Field Trials for the MWA--LFD}

\shortauthors{The MWA--LFD Prototype Deployment Team}

\usepackage{amssymb}
\usepackage{amsmath}
\usepackage{amsbsy}
\usepackage{epsfig}

\begin{document}


\title{Field Deployment of Prototype Antenna Tiles for the Mileura Widefield
Array--Low Frequency Demonstrator}


\author{
 Judd D. Bowman,\altaffilmark{1}
 David G. Barnes,\altaffilmark{4}
 Frank H. Briggs,\altaffilmark{3,9}
 Brian E. Corey,\altaffilmark{2}
 Merv J. Lynch,\altaffilmark{5}
 N. D. Ramesh Bhat,\altaffilmark{7}
 Roger J. Cappallo,\altaffilmark{2}
 Sheperd S. Doeleman,\altaffilmark{2}
 Brian J. Fanous,\altaffilmark{2}
 David Herne,\altaffilmark{5}
 Jacqueline N. Hewitt,\altaffilmark{1}
 Chris Johnston,\altaffilmark{4}
 Justin C. Kasper,\altaffilmark{1}
 Jonathon Kocz,\altaffilmark{3}
 Eric Kratzenberg,\altaffilmark{2}
 Colin J. Lonsdale,\altaffilmark{2}
 Miguel F. Morales,\altaffilmark{1}
 Divya Oberoi,\altaffilmark{2}
 Joseph E. Salah,\altaffilmark{2}
 Bruce Stansby,\altaffilmark{5}
 Jamie Stevens,\altaffilmark{6}
 Glen Torr,\altaffilmark{8}
 Randall Wayth,\altaffilmark{4}
 Rachel L. Webster,\altaffilmark{4}
 J. Stuart B. Wyithe\altaffilmark{4}
 }

\altaffiltext{1}{Massachusetts Institute of Technology, Kavli
Institute for Astrophysics and Space Research, Cambridge,
Massachusetts, USA}

\altaffiltext{2}{Massachusetts Institute of Technology, Haystack
Observatory, Westford, Massachusetts, USA}

\altaffiltext{3}{The Australian National University, Research School
of Astronomy and Astrophysics, Mount Stromlo Observatory, Cotter
Road, Weston Creek, ACT 2611, Australia}

\altaffiltext{4}{School of Physics, University of Melbourne, VIC,
3010, Australia}

\altaffiltext{5}{Department of Applied Physics, Curtin University of
Technology, GPO Box U1987, Perth, Western Australia}

\altaffiltext{6}{Mathematics and Physics, University of Tasmania,
Hobart, Tasmania 7005, Australia}

\altaffiltext{7}{Centre for Astrophysics and Supercomputing,
Swinburne University of Technology, Hawthorn, Victoria 3122,
Australia}

\altaffiltext{8}{The Australian National University, Faculty of
Science, Canberra, ACT 0200, Australia}

\altaffiltext{9}{Australia Telescope National Facility, Commonwealth
Scientific and Industrial Research Organization, PO Box 76, Epping,
NSW 1710, Australia}


\begin{abstract}

Experiments were performed with prototype antenna tiles for the
Mileura Widefield Array--Low Frequency Demonstrator (MWA--LFD) to
better understand the widefield, wideband properties of their design
and to characterize the radio frequency interference (RFI) between 80
and 300 MHz at the site in Western Australia. Observations acquired
during the six month deployment confirmed the predicted sensitivity
of the antennas, sky-noise dominated system temperatures, and
phase-coherent interferometric measurements. The radio spectrum is
remarkably free of strong terrestrial signals, with the exception of
two narrow frequency bands allocated to satellite downlinks and rare
bursts due to ground-based transmissions being scattered from
aircraft and meteor trails. Results indicate the potential of the
MWA--LFD to make significant achievements in its three key science
objectives: epoch of reionziation science, heliospheric science, and
radio transient detection.

\end{abstract}



\section{INTRODUCTION} \label{sec_intro}

With the advent of high-performance, low-cost digital signal
processing capabilities, a new approach to radio astronomy
instrumentation is possible.  At frequencies of a few hundred MHz and
below, it has become feasible to directly sample radio-frequency
waveforms from antennas and perform traditionally analog functions,
such as filtering and mixing, in the digital domain.  This minimizes
analog devices in the signal path and leads to stable and
well-calibrated systems.  The low cost of such approaches opens up
the possibility of deploying a very large number of small antennas,
each equipped with direct-sampling digital receivers, thereby gaining
access to seamless, wide fields of view.

Multiple arrays exploiting this new capability are currently under
development. These instruments include the Mileura Widefield
Array--Low Frequency
Demonstrator\footnote{http://haystack.mit.edu/ast/arrays/mwa}
(MWA--LFD) in western Australia, the Low Frequency
Array\footnote{http://www.lofar.org} (LOFAR) in the Netherlands, the
Primeval Structure Telescope\footnote{\citet{Pen_PAST}} (PAST) in
northern China, and the Long Wavelength
Array\footnote{http://lwa.unm.edu/} (LWA) in New Mexico.

In this paper, we report on a field prototyping effort for the
MWA--LFD. This instrument is designed to cover the frequency range
between 80 to 300 MHz and will be built at Mileura Station in Western
Australia. The site of the MWA--LFD has been chosen as the Australian
candidate site for the international Square Kilometer Array (SKA) and
efforts are well advanced to establish a radio astronomy park in the
region with comprehensive and permanent radio frequency interference
(RFI) control measures.

The primary scientific goals of the MWA--LFD are the characterization
of redshifted 21 cm HI emission from the cosmological epoch of
reionization (EOR) \citep{Furlanetto_EOR_Review, Bowman_Forecast,
2006ApJ...638...20B, McQuinn_Cosmology, 2005ApJ...634..715W,
2005ApJ...635....1B, 2004ApJ...615....7M, 2004ApJ...608..622Z}, a
survey of the sky for astronomical radio transient sources, and the
investigation of the heliosphere through scintillation and Faraday
rotation effects in order to demonstrate the ability of such
measurements to improve the prediction of space weather
\citep{2005SPIE.5901..124S}. All of these applications demand very
wide fields of view, and as a result, the array design features a
large number of small, phased-array antenna tiles with (half maximum)
fields of view of $\sim 200 \lambda^2$ deg$^2$, where $\lambda$ is
the observing wavelength in meters and is in the range $1 \lesssim
\lambda \lesssim3$ m.

The full array will consist of 500 such antenna tiles spread over an
area 1.5 km in diameter and will have a total collecting area of
order $8000$ m$^2$ per polarization. Digital receiver units deployed
across the array will channelize the sampled data streams from each
antenna tile and transmit a selected 32 MHz of sky bandwidth to the
central facility at high spectral resolution. At the central
facility, all antenna pairs will be cross-correlated in order to
avoid multiple layers of beamforming and the consequent boundaries in
sky coverage. Application-specific real-time post-correlation
processing will be implemented in order to sustain the formidable
data rates that will be generated.

As part of the development effort for the MWA--LFD, prototypes of the
antenna tiles were constructed.  In order to test these prototypes,
as well as to characterize the site for RFI and a variety of
environmental factors, we deployed three antenna tiles plus a simple
data capture and software correlation system, along with needed
infrastructure, at the Mileura site.  Four expeditions to the site
were conducted from March through September, 2005.  During this
6-month period, the equipment was operated for a total of 8 weeks and
several terabytes of data were gathered. This program was supported
by MIT, the University of Melbourne, the Australian National
University, and Curtin University of Technology.

In this paper we describe the prototype system and the results
derived from it. In Sections \ref{s_inst} and \ref{s_calib} we
describe the equipment and the physical deployment at the Mileura
site, along with the experimental results relevant to
characterization of the performance of the system. Section
\ref{s_site} addresses site characterization, with particular
emphasis on the RFI environment as seen by our systems. In Section
\ref{s_obs}, a variety of results are presented showing the
characterization of astronomical sources with the prototype
equipment, and our conclusions are given in Section \ref{s_concl}.
Two additional astronomical results of independent scientific value
regarding Type-III solar bursts and giant pulses from the Crab Nebula
pulsar were obtained during the observing campaigns and will be
reported separately.

\section{INSTRUMENT DESIGN}
\label{s_inst}

\begin{figure}
\noindent\includegraphics[width=19pc]{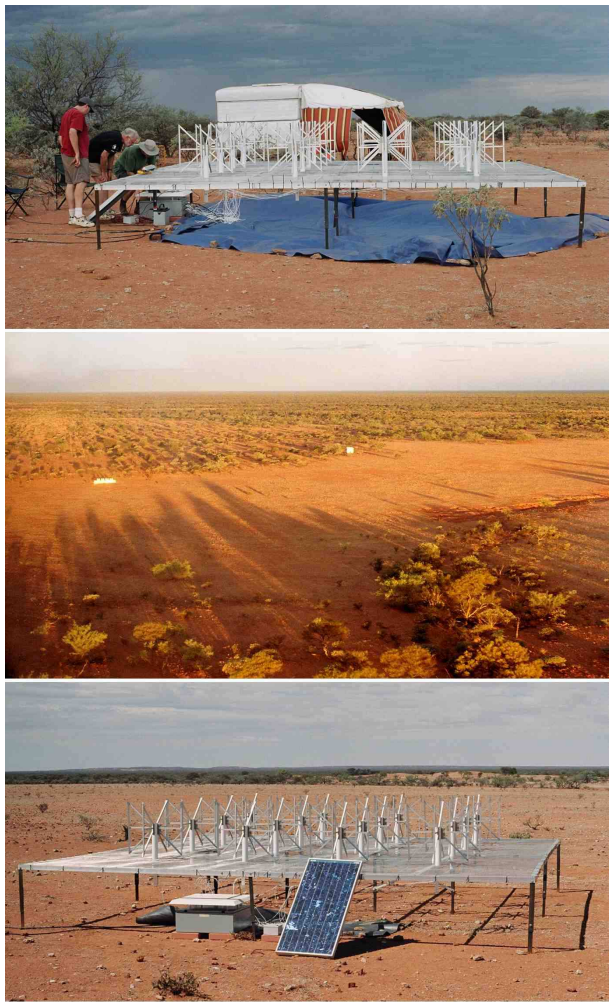} \caption{
\label{f_deployment} Photographs of the first MWA--LFD prototype
antenna tile being tested on site in Mileura, Western Australia. The
antenna consists of 16 crossed-dipoles in a four-by-four meter grid
and is elevated approximately 0.8 m above the ground.  In the top
panel, the antenna tile is being assembled near the caravan used for
housing the computers and receiver electronics. The middle panel is
an aerial view after the antenna tile was moved to its final location
approximately 150 m northwest of the caravan. The bottom panel shows
the completed antenna tile in position.  The solar panel is visible
leaning against the tile and the beamformer box and heat shield are
under the tile.  Two additional antenna tiles were subsequently
installed at the site. }
\end{figure}

The basic antenna concept is shown in Figure \ref{f_deployment} and
is a four-by-four phased-array of wideband active dipoles forming an
``antenna tile" operating in the 80 to 300 MHz frequency range. The
antenna tile is electronically steerable and, due to the wide
frequency range, a switched delay-line beamformer design is employed.
The spacing of the dipoles in the four-by-four grid is 1.07 meters,
or $\lambda / 2$ at 140 MHz.  This choice of spacing is aimed at
optimizing performance for EOR science, but leads to significant
grating lobes at higher frequencies.

The dipoles are a vertical bowtie design, with symmetry above and
below the midline in order to minimize horizon gain (see Figure
\ref{f_dipole}). An active balun is located at the junction of the
bowtie arms, and the amplified signal is fed to the beamformer unit
via coaxial cable. For the prototype units, structural support for
these dipole elements was provided by a Lexan tube of 6.25 cm
diameter.

Because of imperfect impedance matching between the dipole and the
balun, much of the antenna power received from the sky is reflected
back to the sky, particularly at the low end of the frequency range.
The sky brightness temperature is so much higher than the first-stage
amplifier noise temperature, however, that the overall system
temperature is dominated by sky noise, despite the loss of the
reflected power (see Section \ref{s_temp} for additional discussion
of the system temperature).

The dipoles need to be positioned over a ground plane, and due to the
Lexan tube support structure of the prototypes, it was necessary to
construct the ground planes as raised platforms, with a tube clamping
arrangement below the conductive screen.  This arrangement is visible
in Figure \ref{f_deployment}. Ground planes for each of the three
antenna tiles were custom-built, with successive refinements for each
ground plane based on experience from the previous antenna tile. For
the full array, we expect to use conductive mesh at ground level.

The beamformer units are based on switchable delay lines implemented
as coplanar waveguides via traces on printed circuit boards.  Switch
settings are controlled by a computer workstation communicating with
a microcontroller over an RS-232 serial line.  Certain beamformer
functions, such as power combination, were implemented in the
prototype units with commercially built, connectorized modules, and
the general degree of component integration was lower than that to be
implemented in the full array.  However, the prototype beamformer is
expected to match the production version closely in performance.

\subsection{Wide-band Active Dipole Elements}

\begin{figure}
\noindent\includegraphics[width=19pc]{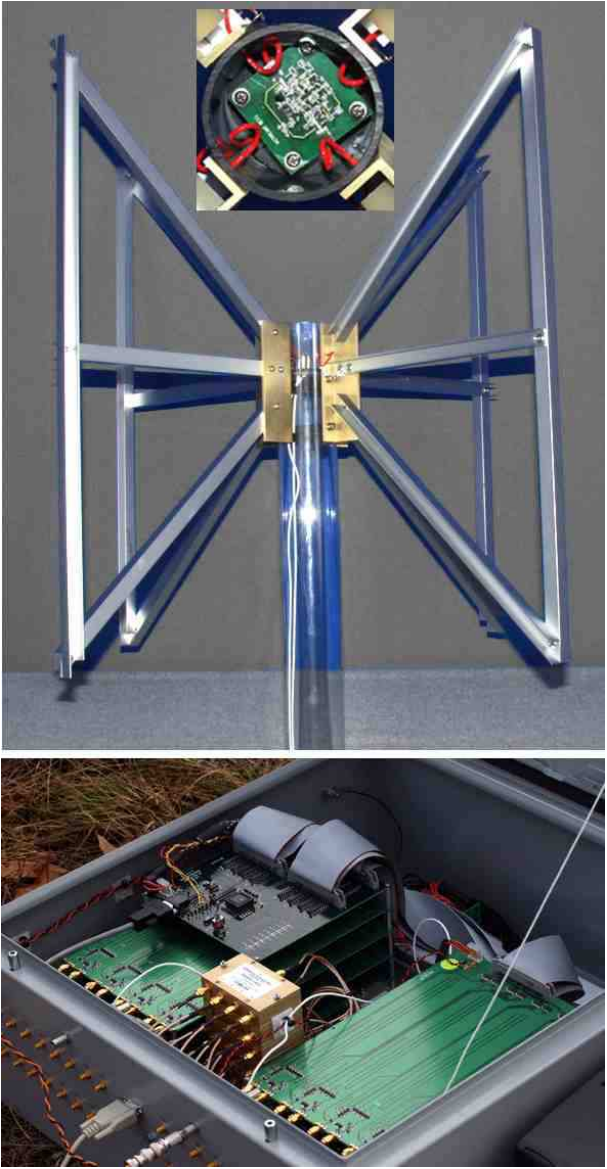} \caption{
\label{f_dipole} Individual bowtie dipole assembly (top) with
close-up of the low-noise amplifier and balun (inset), and interior
of a beamformer (bottom). The signals from sixteen dipole assemblies
are combined by the beamformer to create the output for a single,
phased-array antenna tile. }
\end{figure}

\begin{figure}
\noindent\includegraphics[width=19pc]{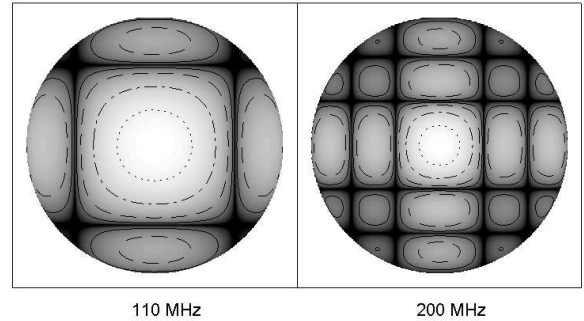} \caption{
\label{f_beampatterns} Predicted beam patterns at two frequencies for
antenna tiles phased to point at the zenith. The left panel shows the
beam pattern at 110 MHz and the right panel at 200 MHz.  The gray
scale is linear in dB and, from black to white, spans -50 to 0 dB.
The contour lines are at -30 dB (solid), -20 dB (dash), -10 dB
(dash-dot), and -3 dB (dot).  The plots are polar projections
spanning the full sky. The predicted beam patterns were derived using
a simple model based on canonical dipoles placed one-quarter
wavelength above an infinite ground plane. No coupling between the
dipole elements was included.}
\end{figure}

\begin{figure}
\noindent\includegraphics[width=19pc]{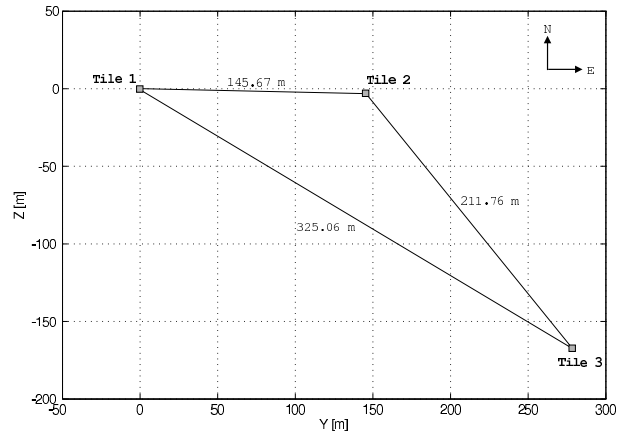} \caption{
\label{f_layout} Diagram of the relative antenna tile locations. The
coordinate frame is a topocentric system and is described in Section
\ref{s_baseline}. The caravan control center was located
approximately in the the center of the array.}
\end{figure}

\begin{figure*}
\noindent\includegraphics[width=39pc]{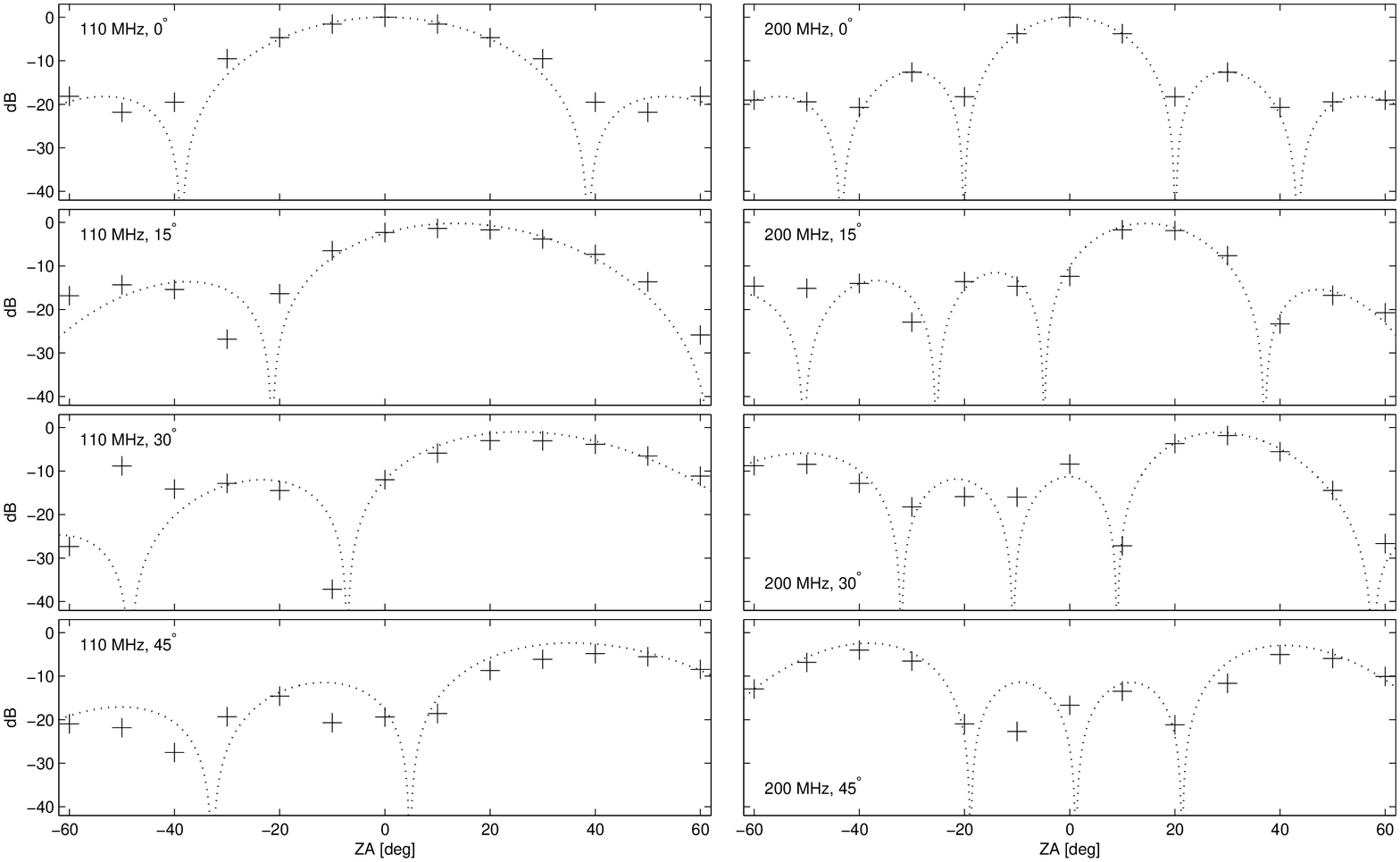} \caption{
\label{f_beammeasurements} Measured power response profiles in the
E-plane of an antenna tile.  The left panels show the measured
profiles at 110 MHz and the right panels at 200 MHz.  From top to
bottom the panels show different antenna tile pointing angles, and
are 0, 15, 30, and 45$^\circ$ relative to the zenith in the E-plane.
In both columns, the dotted lines show the predicted profiles
assuming ideal dipoles and no mutual coupling. The grating lobes of
the antenna tiles are especially evident in the measured data for the
45$^\circ$ pointing angle at 200 MHz.}
\end{figure*}

\begin{figure}
\noindent\includegraphics[width=19pc]{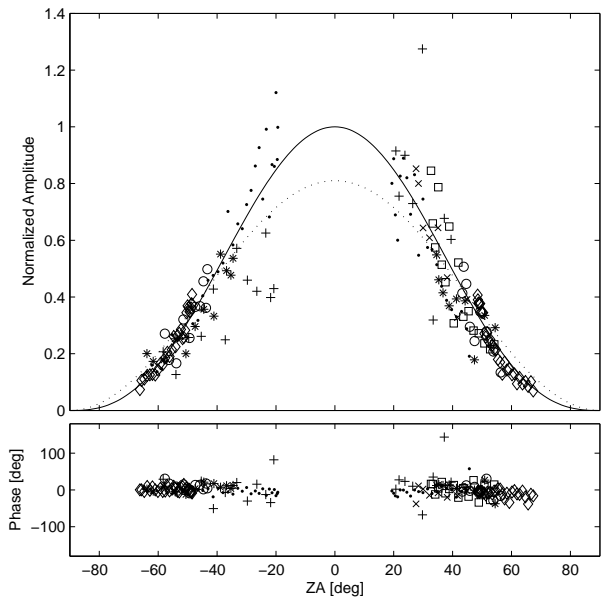} \caption{
\label{f_dipoleshape} Best-fit dipole power envelope in the
north-south polarization at 120 MHz from the baseline formed by
antenna tiles 1 and 3. Seven point sources were isolated
interferometrically and tracked across the sky to estimate the
profile. The sources were PKS2152-69 (circle), PKS2356-61 (star), Pic
A (dot), 3C161 (plus), 3C353 (cross), Her A (square), and Tau A
(diamond). The amplitudes of the resulting visibility measurements
were normalized and fit to a $cos^{\beta}(\theta)$ profile, where
$\theta$ is the zenith angle and the best-fit (solid line) was found
for $\beta=2.6\pm0.4$.  Negative zenith angles indicate measurements
where the source was in the eastern half of the sky, and positive
zenith angles indicate the western half.  The dotted line shows the
equivalently normalized ideal dipole ($\beta=2$) envelope. }
\end{figure}

\begin{figure*}
\noindent\includegraphics[width=39pc]{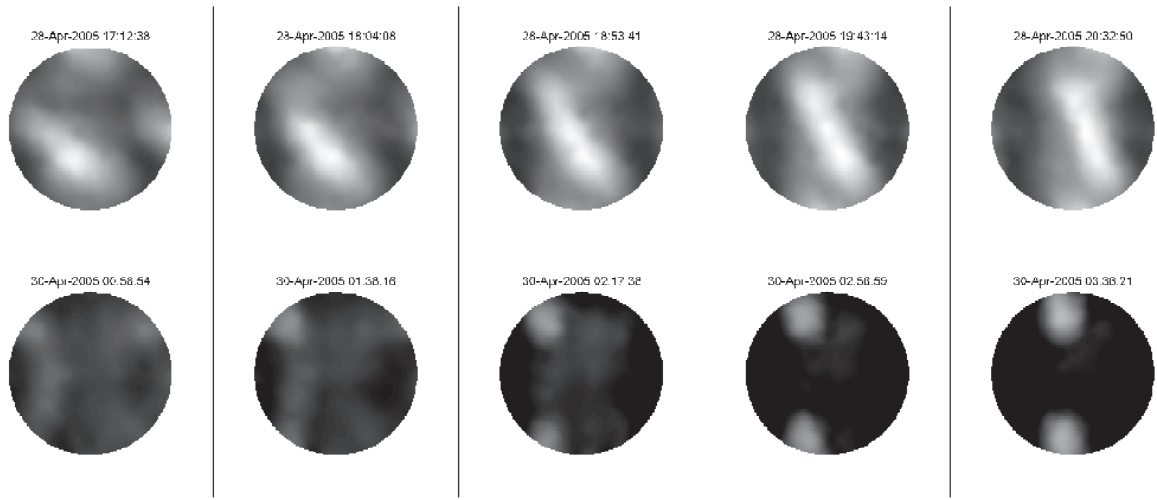} \caption{
\label{f_skymap} Time series of sky maps at 200 MHz produced by
scanning repeatedly an antenna tile beam through a raster of pointing
directions. The Galaxy center is shown rising and transiting in the
top row, and the sun is rising in the bottom row.  The maps are polar
projections from 0 to 60 deg zenith angle with north at the top, and
the shading indicates the sky temperature on a log scale spanning
approximately 250 (black) to 1000 K (white). The effect of the
diffraction grating sidelobes in the antenna tile power response
function are evident in the observations of the sun, producing mirror
images at each corner of the map.  The actual image of the sun is
toward the top-left (northeast) corner of the maps, although the
mirror image in the bottom-left (southeast) is nearly as strong. No
corrections were applied to the data used to generate these maps; in
particular, there is no compensation for the reduced gain at large
zenith angle due to the dipole element power envelope. }
\end{figure*}

\begin{figure}
\noindent\includegraphics[width=19pc]{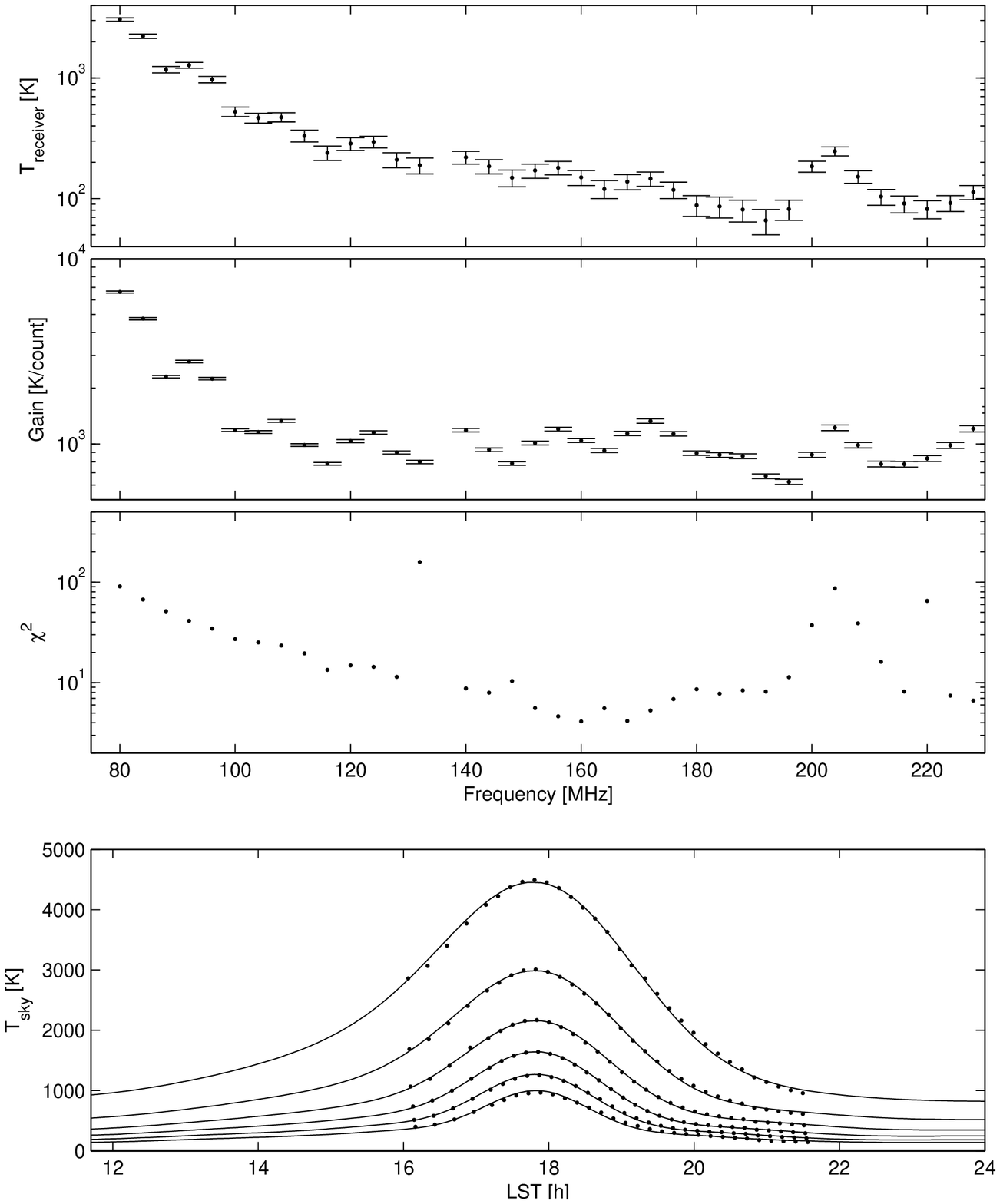} \caption{
\label{f_driftscan} Effective receiver temperature (top) and gain
(second) for 37 frequencies spanning 80 MHz to 228 MHz for the
east-west polarization of antenna tile 1.  Error bars are at 95\%
confidence. The instrumental performance was determined by fitting
Galactic drift scan profiles to observed antenna power measurements.
The third panel indicates the quality of the fit at each frequency by
plotting the $\chi^2$ value. The bottom panel shows example best fits
at 100, 120, 140, 160, 180, and 200 MHz (from top to bottom), where
the solid line is the model and the dots are the data.  Due to an
equipment malfunction affecting the impedance matches within the
beamformer at the time of the observations, these results represent a
lower limit on the system performance. }
\end{figure}

Each dipole element, shown in Figure \ref{f_dipole}, for the
phased-array antenna tile is an ``active antenna," with a low-noise
amplifier and a balun integrated into the antenna structure. The
combination provides low-noise amplification at the point where the
received signal is weakest, and converts it from a balanced to an
unbalanced signal, as is necessary before transmitting the signal
over coaxial cable to the downstream electronics.

The open-frame, vertical bowtie construction of the antenna elements
was chosen with several objectives in mind: wide frequency range,
broad angular coverage, low sensitivity at the horizon (to minimize
reception of RFI from terrestrial sources), and low manufacturing
cost. A simple, horizontal, wire dipole over a ground plane satisfies
the latter two, but has difficulties with the first two---especially
with a two octave frequency range.  The impedance of a dipole has a
much stronger frequency dependence than does the input impedance of a
typical amplifier.  As a result, there is a large impedance mismatch
between dipole and amplifier except over a narrow frequency range,
and most of the power received from the sky is lost back to the sky.
Extending the dipole arms vertically into bowties significantly
flattens the frequency dependence of the antenna impedance, thereby
improving the match with the amplifier over most of the 80 to 300 MHz
range.  The conversion to bowties also widens the beamwidth.
Constructing the bowtie in outline form, with spars around the outer
edges and a single horizontal crosspiece, gives performance similar
to a solid-panel bowtie, but with the advantages of reduced wind
loading and lighter weight, both of which allow a simplified
mechanical support structure and reduce the manufacturing cost.

Each pair of diametrically opposite dipole arms feeds a low-noise
amplifier housed in the Lexan tube.  The amplifier employs two
Agilent ATF-54143 HEMT amplifiers in a balanced configuration with a
balun on the output.  The amplifier noise temperature is 15 to 20 K
(compared with approximately 100 to 5000 K for the sky temperature).
The balanced design and intrinsically low distortion characteristics
of the HEMT ensure an acceptably low level for any output
intermodulation products arising from RFI signals received by the
antenna.

\subsection{Antenna Tile and Delay-Line Beamformer}
\label{s_beamformer}

For each of the 16 signals of the same polarization from the dipole
elements of an antenna tile, the beamformer filters the signal to
attenuate out-of-band interference, and then applies a delay between
0 and 10.4 ns, as appropriate to steer the phased-array beam in the
desired direction.  The delay maximum of 10.4 ns is sufficient to
phase the beam to the horizon in the principle planes of the tile,
but corner dipoles may be under-delayed in the case of off-axis
incident waves with zenith angles greater than 45$^\circ$. At a
zenith angle of 60$^\circ$ and azimuth of 45$^\circ$, the corner
dipole opposite the reference has a serious delay error of 2.7 ns
(0.81 turn at 300 MHz), while the next worst delay errors are only
0.5 ns (0.15 turn at 300 MHz).

The delay for each dipole is generated from five delay lines that may
be independently switched in or out of the signal path via paired
GaAs switches.  The delays differ by powers of two between sections,
so the minimum delay step is ~0.34 ns, or ~0.1 turn at 300 MHz.  The
coplanar-waveguide construction of the delay lines affords excellent
delay repeatability between units and extremely low temperature
sensitivity. Following the delay line sections, the signal is
combined with the other 15 signals of the same polarization, further
amplified, and then sent on to the receiver via coaxial cable.

Each beamformer chassis, shown in Figure \ref{f_dipole}, contains the
delay lines and associated electronics for the two polarizations from
a single tile, voltage regulators to step down the DC voltage from an
external power source, and a microcontroller-based digital interface
to convert RS-232 serial data commands into parallel data. These
signals control the delay line switches and also a second set of
switches, one per dipole signal, which allow individual dipoles to be
switched in or out of the signal path for test purposes or in the
event of a dipole failure.  During astronomical observations, the
control electronics are deactivated to reduce interference except for
the brief instants when the delays are reset to change the pointing
of the antenna tile beam.

The mutual coupling between antenna elements is less than $-20$ dB,
except between 90 and 140 MHz, where it peaks at about $-12$ dB at
110 MHz. With this low mutual coupling between antenna elements, the
antenna tile power response pattern is modelled as the single-element
dipole pattern multiplied by the array factor, which is the pattern
the antenna tile would have if all the elements were replaced by
ideal, isotropic antennas. The half-power beamwidth of a single
dipole element mounted above a ground plane is $\gtrsim 80$ deg over
most of the frequency range, with peak response at the zenith except
near 300 MHz.  The beamwidth of a full antenna tile varies
approximately inversely with frequency, and has zenith values of 15
to 45 deg. The beamwidth increases as the beam is steered down toward
the horizon due to foreshortening. Figure \ref{f_beampatterns} shows
two example antenna tile power response patterns calculated assuming
ideal dipole elements with no coupling.  The small actual coupling
does cause slight deviations from these patterns, as discussed in
Section $\ref{s_ant_pow}$.

\subsection{Receiver and Digitizer System}

For the prototype field deployment, the receiver functions were
handled by a simple, dual-frequency conversion system, wherein a 4
MHz band was selected and mixed to 28 MHz center frequency.  The
signal was digitally sampled at 64 MHz using the ``Stromlo Streamer"
\citep{Briggs_Streamer}, but only one in four samples was kept,
producing an effective sampling rate of 16 MHz.

The system used a computer workstation equipped with a terabyte RAID
array to record four input channels at the effective sampling rate
with 8 bit precision, giving a data rate of 64 MB/s that was recorded
to disk. Normally, only the low order bits were toggled by the
received signals. The dynamic range of the 8 bits allowed operation
of the system over the full frequency range without further gain
adjustment. There was ample head-room for RFI signals, and in those
few instances where higher order bits were needed, the IF conversion
system entered a non-linear regime due to the high signal power
concentrated in the 4 MHz passband at the output of the IF system.
The data were processed in software by playing the recordings back
into an FX correlator code which computed both auto- and
cross-correlation power spectra. There were two principal modes of
operation: a high spectral resolution mode, which implemented a
polyphase filter bank to obtain 1 kHz spectral channels with a high
level of channel-to-channel rejection, and a 16 kHz resolution mode,
which used a straight FFT for spectral dispersion in order to cut the
correlation time to a little longer than recording time.  This latter
mode produced 512 channels and was used in the analysis in Section
\ref{s_solar}.

\subsection{Array Layout}

The location of the prototype deployment array is given in Table
\ref{t_location}.  At the site, the three antenna tiles were deployed
in a triangular configuration roughly 300 m across, as illustrated in
Figure \ref{f_layout}. This provided interferometer baselines of
sufficient length to resolve out the diffuse galactic emission and to
effectively isolate signals from bright individual point sources
through phase-stable interferometry. The configuration also allowed
the potential of the array to yield scientifically interesting
astronomical information.

\subsection{Site Infrastructure}

The prototype field deployment campaigns had modest infrastructure
requirements. A small caravan (trailer) was purchased and placed
on-site to act as the control center and house the receivers and
computer equipment. Battery-backed solar power units were installed
to power each of the three antenna tiles and a portable diesel
generator was used at the caravan to provide power for the
electronics and daytime air conditioning. Coaxial cables of 200 m
length were used to transport signals from the antenna tile
beamformers to the caravan, and the cables were simply laid on the
ground.  Between campaigns, the beamformers and solar power units
were removed and placed in storage.

\section{CALIBRATION AND PERFORMANCE}
\label{s_calib}

\subsection{Antenna Power Response Pattern}
\label{s_ant_pow}

The actual antenna power response pattern for the prototype antenna
tiles was constrained using two methods. Prior to installing the
antenna tiles in the field, the response function was determined at
two frequencies by systematically scanning a transmitter above an
antenna tile using facilities at the MIT Haystack Observatory. Figure
\ref{f_beammeasurements} shows resulting beam profiles for line scans
in the E-plane of the antenna tile at four different antenna tile
pointing angles.  The structures in these profiles agree well with
the patterns produced by assuming that the antenna tiles consist of
ideal dipole elements above an infinite ground plane with no mutual
coupling, which are shown as the dotted lines in Figure
\ref{f_beammeasurements}.  There are a few notable deviations.  Most
prominently, the primary beam appears larger than predicted at 110
MHz, as can be seen in the upper-left two panels in the Figure.
Additionally, the nulls in the beam patterns are not as deep as would
be expected from the ideal calculations, and at the larger pointing
angles, there are some significant deviations from the expected
patterns.  Some of these discrepancies may be the result of the test
conditions.  The distance ($\sim$15 m) between the transmitter and
antenna tile was not sufficient to reach fully the far-field limit
for the radiation patterns, and the alignment and pointing of the
transmitter at the antenna tile was performed manually with an
estimated margin of error of $\sim5^\circ$ in each, allowing
ambiguities in the effective zenith angle and in gain due to the
dipole pattern of the transmitter.

In the field, additional measurements were achieved by
interferometrically isolating point sources and tracking them as they
drifted across the sky. This technique effectively separated the
dipole element power response pattern from that of the full antenna
tile pattern and resulted in an estimate of the effective dipole
power envelope as a function of zenith angle. The amplitudes of the
interferometrically isolated point sources were fit to a
$cos^{\beta}(\theta)$ profile, where $\theta$ is the zenith angle and
$\beta$ is a constant. The best-fit at 120 MHz is shown in Figure
\ref{f_dipoleshape} and was found to be given by $\beta=2.6\pm0.4$,
whereas an ideal dipole has $\beta=2$ and is excluded by the
measurements at the 85\% confidence level.  This indicates that the
effective power pattern for the antenna tile dipole elements has a
somewhat narrower shape than the ideal case.

\subsection{Gains and System Temperatures}
\label{s_temp}

The gain and noise temperature of an antenna tile system was measured
by setting its beamformer delays to point the antenna tile towards
zenith and then recording the signal over a long period of time as
the Earth turns and the beam transits the Galactic plane. In this
technique, the measured power is compared to the predicted power by
convolving a map of the low frequency Galactic synchrotron emission
with the computed antenna power response function. The data are fit
by \cite[Their Equation 1]{2004RaSc...39.2023R},
\begin{equation}
P(t) = g  \left [ T_{sky}(t, \boldsymbol \theta) \star W(\boldsymbol
\theta) + T_{receiver} \right ],
\end{equation}
where $P$ is the measured power in arbitrary units as a function of
local sidereal time (LST), $g$ is the receiver gain, $T_{sky}$ is the
sky model as function of direction in the sky $\theta$, $W$ is the
antenna power response pattern, $T_{receiver}$ includes the remaining
system noise contributions (including amplifier noise and ground
spillover) referred to the antenna output, and $\star$ is the
convolution operator.

We used the 408 MHz sky map made by \citet{1982A&AS...47....1H} and
assumed a spectral index of $\alpha=2.6$ to extrapolate to the
observed frequencies.  The power response pattern for the prototype
antenna tiles was taken to be that of ideal dipole elements with no
mutual coupling above an infinite ground plane, as described at the
end of Section \ref{s_beamformer}. Although this simplified model was
shown in Section \ref{s_ant_pow} to produce a larger dipole power
envelope than that of the actual antennas tiles, it deviates by only
approximately 10\% at the edge of the primary antenna tile beam when
phased to the zenith.

Figure \ref{f_skymap} gives examples of the actual sky as seen by a
prototype antenna tile, and Figure \ref{f_driftscan} shows the
results of the system gain and noise temperature calibration between
80 and 230 MHz. In general, the effective receiver temperature as
measured at the output of the antenna was found to be below about 200
K except at the lowest frequencies measured, although, the results
are considered a lower limit on the performance of the system due to
an equipment problem at the time of the observations which increased
impedance mismatches within the beamformer of the antenna tile used
for the observations. The predicted antenna temperature as a function
of LST and examples of the best least-squares fits of the data at six
frequencies are also given.

\subsection{Baseline Determination}
\label{s_baseline}

Baselines of the final array configuration were determined using
observations of unresolved astronomical continuum sources. The
antenna coordinates were defined in the conventional right-handed,
cartesian coordinate system with the $x$- and $y$-axes lying in a
plane parallel to Earth's equator and the $z$-axis parallel to
Earth's rotation axis pointing towards the north pole. In terms of
hour angle (HA) and declination (Dec), $x$, $y$, and $z$ are measured
towards (HA = 0h, Dec = 0$^{\circ}$), (HA = -6h, Dec = 0$^{\circ}$)
and (Dec = 90$^{\circ}$) respectively. In this coordinate system, a
large HA coverage is useful to constrain the $x$ and $y$ components
of the baseline, and the $z$ component is best constrained by
observing sources spanning a large range in declination.

An observing session optimized for baseline calibration was conducted
on the night of September 15, 2005 which continued through
mid-morning the next day. The observations were carried out at 120
and 275 MHz using the north-south aligned polarization on all three
tiles and observed seven bright, unresolved sources: Tau-A, Her-A,
3C252, 3C161, Pic-A, PKS 2152-69 and PKS 2356-61. The relative
antenna tile locations are given in Table \ref{t_tiles}, and Table
\ref{t_comparison} shows a comparison between the baseline vectors
determined by astronomical observations and those obtained manually
by using a tape measure.

The resulting antenna tile positions are sufficiently accurate to
correct the interferometer phases for changing geometrical delays as
a function of hour angle and declination. The least square fitting
uncertainties in $x$ and $y$ were determined to be about 2 cm and
those in $z$ to be about 5 cm. The dominant sources of error in the
antenna tile coordinates due to pseudo-random effects were cross talk
in the receiver chains (see Section \ref{s_cross_talk}) and
ionospheric variations (see Section \ref{s_obs_ionosphere}).

\subsection{Antenna Delays}

In addition to determination of the antenna coordinates, calibration
of interferometric data requires the determination of antenna based
delays (or phase offsets) due to differences in the electrical path
lengths to different antennas. These offsets were determined using
observations of Pic-A conducted on September 18, 2005. An analysis
using all three baselines sensing the east-west aligned polarization
led to the conclusion that the electrical paths to antenna tiles 1
and 2 differ by 1.5 m and that those to antenna tiles 1 and 3 differ
by 0.6 m. The major contributor to these difference is differing
lengths of the cables between the antennas and the inputs to the
digitizers. The measured instrumental phase offsets were found to be
well behaved, varying with frequency $f$ as $\Delta \phi = 2\pi
\Delta L f/c$ where $\Delta L$ is the electrical path difference and
$c$ is the speed of light.

\subsection{Cross Talk}
\label{s_cross_talk}

When the array was pointed at a bright radio source, the
interferometers showed a response with two components: one
corresponding to the natural fringe rate due to the Earth's rotation,
and a second that was invariant with time. Due to the simplicity of
the receiving system, it was deduced that the invariant component is
dominated by an instrumental component, caused by cross talk (faint
electrical coupling) between the signal paths in the receiving
system. In more sophisticated receivers, this effect can be avoided
by phase switching the signal close to the antenna, followed by
synchronous demodulation in the digital section, and the full array
will employ this technique.

For the purposes of this prototyping exercise, it was sufficient to
approximate the invariant term by a long average of the
interferometer response at ``zero fringe rate'' (i.e., computed
without compensation for Earth rotation) and to remove the cross talk
component by subtraction of the average from the visibility data
prior to application of the fringe de-rotation for the celestial
sources.

\subsection{Noise and Integration Time}
\label{s_noise_vs_time}

\begin{figure}
\noindent\includegraphics[width=19pc]{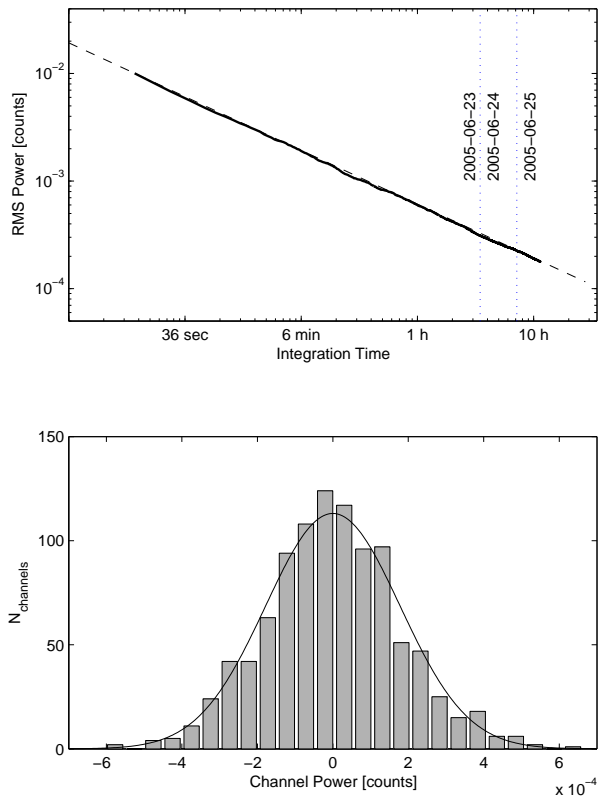} \caption{
\label{f_rms_vs_time} Analysis of the 10 h integration centered at
187 MHz. The top panel shows the RMS channel power for the 1000
frequency channels spanning 188 to 189 MHz as a function of
integration time and clearly follows a $t^{-1/2}$ trend (shown by the
dashed line). The lower panel gives the distribution of channel
powers at the end of the integration, which is close to an ideal
Gaussian distribution (shown by the solid line).}
\end{figure}

In order to assess the noise characteristics of the system, a deep
integration was acquired in an empty part of the spectrum with one
antenna tile.  The observation was conducted over three days
beginning June 23, 2005, and resulted in 10 h of total integration in
a 4 MHz band centered at 187 MHz with 1 kHz spectral resolution. The
integration was processed to remove the bandpass and corrected for
amplitude variations due to variations in Galactic noise with LST.
From these cleaned data, 1000 contiguous channels spanning 188 to 189
MHz were selected and analyzed as a function of the cumulative
integration time. For pure thermal noise, the variation of the
individual channel powers about the mean should have a Gaussian
distribution and their RMS should be proportional to the inverse
square root of the integration time ($\sim t^{-1/2}$). Figure
\ref{f_rms_vs_time} shows that the observed trend is consistent with
thermal noise and that the distribution of channel powers at the end
of the integration is approximately Gaussian.

\section{SITE CHARACTERIZATION}
\label{s_site}

The Mileura site is a new venue for radio astronomy and the prototype
field deployment system was one of the first instruments to be
deployed at the site. The effort, therefore, was able to provide
valuable experiences regarding working conditions in the remote area,
the ability of the equipment to withstand the harsh environment, and
the suitability of the site for low frequency radio astronomy.  In
this Section, we report significant findings on these topics.

\subsection{The Mileura Environment}

Mileura Station is an active sheep and cattle station (ranch) in a
remote area approximately 620 km north of Perth, Western Australia.
The nearest towns are Meekathara ($\sim$100 km east) and Cue
($\sim$150 km southeast), and the nearest small city is Geraldton
($\sim$350 km southwest). During dry seasons, Mileura can be reached
from Perth by car or truck in approximately 10 hours over paved and
dirt roads. In wet weather, the dirt roads may become virtually
impassable for short periods of time, cutting off access to the site.
The ranch consists of about 620,000 acres ($\sim$2500 km$^2$) of
natural grazing land for sheep and cattle, but kangaroos, emus, feral
goats, lizards, snakes, birds, and many insects are also abundant.
The climate of the region is arid (see Table \ref{t_climate}).

Prominent concerns prior to the field deployment effort regarded the
logistics of working at the site and whether the equipment would be
able to endure the harsh environment, including the climate and the
wildlife.  Over the course of the six-month deployment, the equipment
was operated successfully in a variety of conditions including
extreme heat, severe winds, cool nights, and light rain.
Additionally, although numerous kangaroos and emus were seen in the
vicinity of the array, no damage or evidence of any interaction with
the equipment was observed.

Several severe storms occurred while the array was not in use. One of
these storms produced flooding at the site.  Inspection of the
antenna tiles after the flooding showed clear signs that running
water passed under the elevated ground planes, eroding material
around the ground plane supports.  A second storm produced large hail
(up to $\sim$3 cm in diameter) at the site, but no significant damage
was detected to the antenna tiles, including the dipole elements and
ground planes.

\subsection{Radio Frequency Interference}

\begin{figure*}
\noindent\includegraphics[width=39pc]{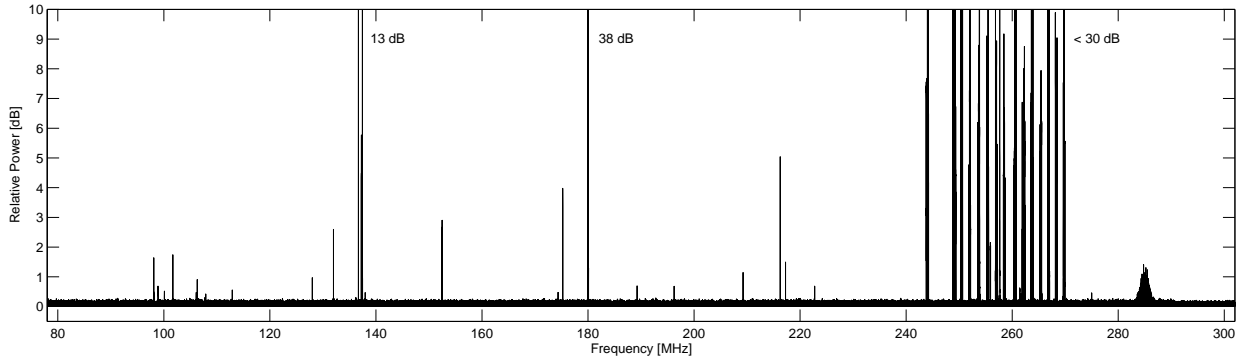} \caption{
\label{f_fullspectrum} Measured RFI spectrum at Mileura from 80 to
300 MHz. The full spectrum was generated from 55 separate 15-second
integrations, each covering 4 MHz. The resulting spectra, with 1 kHz
resolution, were bandpass-corrected and stitched together into one
long spectrum. At the lower frequencies, a low-amplitude ripple of
600 kHz periodicity, caused by cable reflections, was removed by
median filtering. At all frequencies, the 0 dB level corresponds
approximately to the Galactic background noise in a cold part of the
sky. Although the physical resolution on the plot corresponds to
several hundred channels, even a single 1 kHz channel with RFI will
be visible. There are many long sections of spectrum completely free
of RFI at this sensitivity.}
\end{figure*}

Characterizing the RFI at the Mileura site was an important objective
of the prototype field deployment effort and a significant amount of
observing time was dedicated to this task.  All RFI studies were
processed with the high spectral dynamic range, polyphase filter bank
mode to obtain 1 kHz resolution. Figure \ref{f_fullspectrum} gives an
overview of the strongest signals in the 80 to 300 MHz band.  This
full spectrum scan results from the stitching together of a sequence
of 4 MHz bands. Each individual band was observed for 15 seconds by a
single antenna tile phased to point at the zenith and was bandpass
corrected.  With 1 kHz channels, less than 1\% of the spectrum was
occupied by interferers above 5-$\sigma$ in this observation. The
strongest identified features, most notably at 137, 180, and 240
through 270 MHz, originate from satellites. The temporary receiving
system used in these prototyping exercises emits signals at multiples
of the 64 MHz sampling clock that could be detected in the deep
spectra. The full array will use a much higher sampling rate, and the
critical components will be contained in RFI tight enclosures.

\subsubsection{Long Integrations}

\begin{figure}
\noindent\includegraphics[width=19pc]{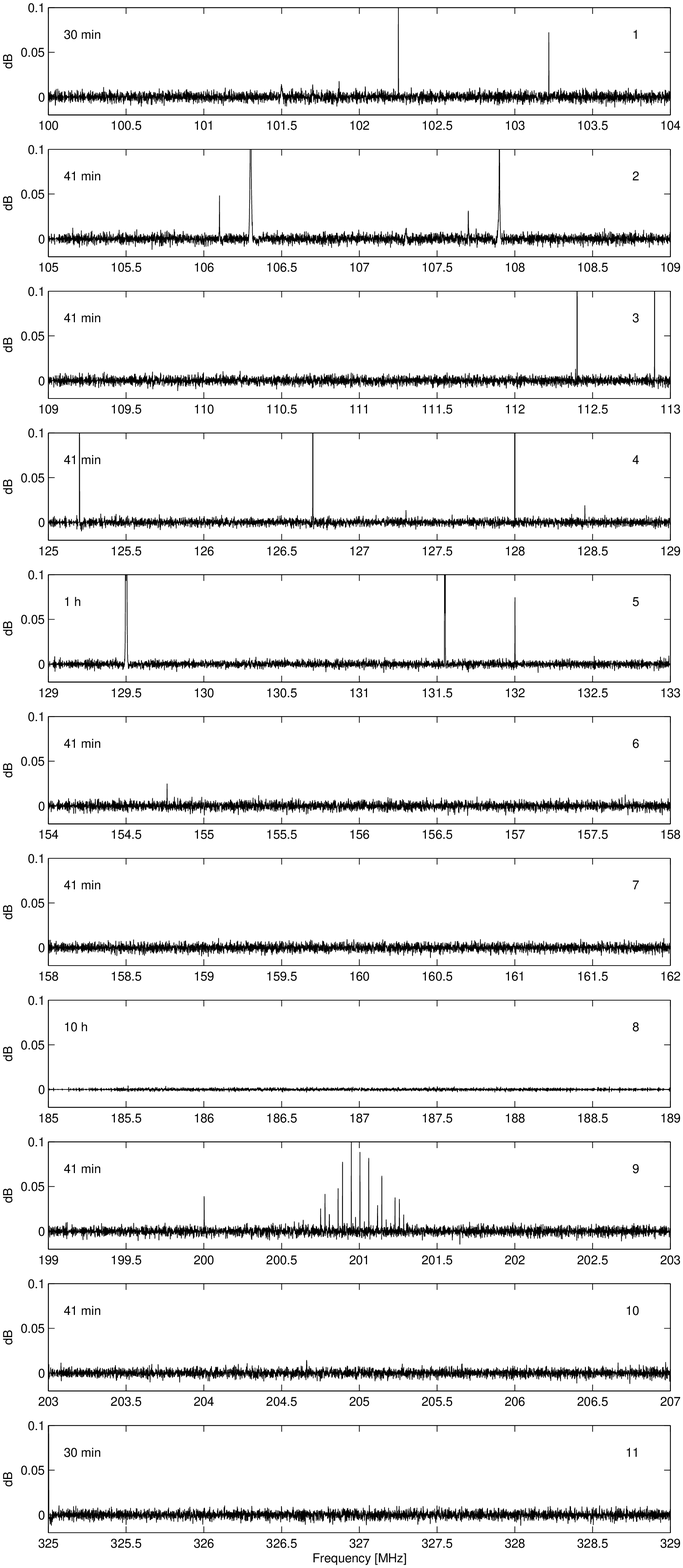} \caption{
\label{f_rfi_deep} Deep integrations with a single antenna tile for
11 spectral windows of 4 MHz each in order to characterize the
low-level RFI environment. Integration times are given in the top
left corner of each plot and range from 30 min to 10 h. The scales
are in dB relative to the system noise, which is dominated by the sky
background. As discussed in Section \ref{s_site}, much of the visible
RFI in these observations is due to intermittent sources. }
\end{figure}

Figure \ref{f_rfi_deep} shows spectra for long integrations in 11
selected bands.  The observations vary between 30 min and 10 h and
each was acquired with a single antenna tile phased to point at the
zenith. The sensitivity levels probed with these deep integrations,
where the time-bandwidth product is $\sim10^6$-$10^7$, achieve RMS
channel powers more than 30 dB below the galactic background at 1 kHz
resolution. These observations illustrate that the spectrum is
remarkably free of RFI.

\begin{figure}
\noindent\includegraphics[width=19pc]{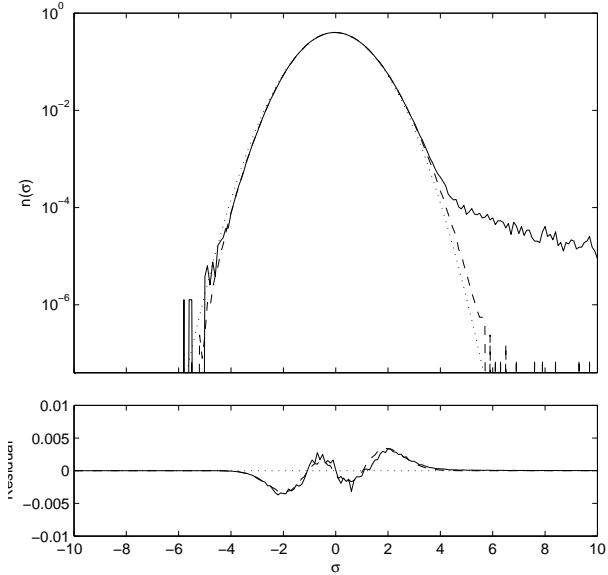} \caption{
\label{f_rfi_hists} Normalized probability of individual channel
power measurements deviating from the mean value for two deep
integrations (top) and the residual variation from purely Gaussian
noise (bottom). The solid lines are for the 42 min integration
centered at 107 MHz, and the dashed lines are for the 10 h
integration centered at 187 MHz.  The dotted lines give a reference
Gaussian distribution.  Each integration consists of 4096 channels
spanning 4 MHz and is divided into 1.3 sec time steps.  Thus, the two
observations have $4096\times1922=7,872,512$ and
$4096\times30874=126,459,904$ individual samples, respectively. The
high-$\sigma$ tail in the 107 MHz data is due to the presence of
RFI.}
\end{figure}

The RFI in a particular observation can be characterized by
considering the histogram of individual 1 kHz by 1.3 sec samples
comprising the full integration.  Absent of RFI and systematic
processing errors, the distribution of samples should approach a
Gaussian distribution.  Figure \ref{f_rfi_hists} shows two such
histograms for the 42 min and 10 h observations centered at 107 and
187 MHz, repsectively (panels 2 and 8 in Figure \ref{f_rfi_deep}).
The significant presence of RFI due to FM radio transmissions in the
observation centered 107 MHz is quite clear as a deviation above a
pure Gaussian distribution at high $\sigma$, while the very clean
spectrum around 187 MHz reproduces a Gaussian distribution more
accurately. In both cases, there appears to be a slight systematic
deviation (shown by the bottom plot in Figure \ref{f_rfi_hists}) from
the best-fit Gaussian at all levels. The deviation is most likely an
artifact of the bandpass correction applied to both observations.

The 10 h integration centered at 187 MHz (also discussed in Section
\ref{s_noise_vs_time}) is a particularly clean observation and has
significant implications for the measurements of redshifted 21 cm HI
emission that will be targeted by EOR experiments with the full
MWA--LFD.  The planned EOR observations will require very deep
integrations, of order 100 to 1000 h, to achieve the required
sensitivity levels. The duration of the observation centered at 187
MHz, therefore, represents a substantial step towards characterizing
the properties of the RFI at levels approaching those of interest,
and the lack of detectable interferers (even if over only a small
portion of the total spectrum) bolsters confidence that the full
MWA--LFD will be able to achieve the very deep integrations needed to
detect evidence of fluctuations in the redshifted 21 cm HI emission
from the EOR.

\subsubsection{Reflections from Meteor Trails and Aircraft}

\begin{figure*}
\noindent\includegraphics[width=39pc]{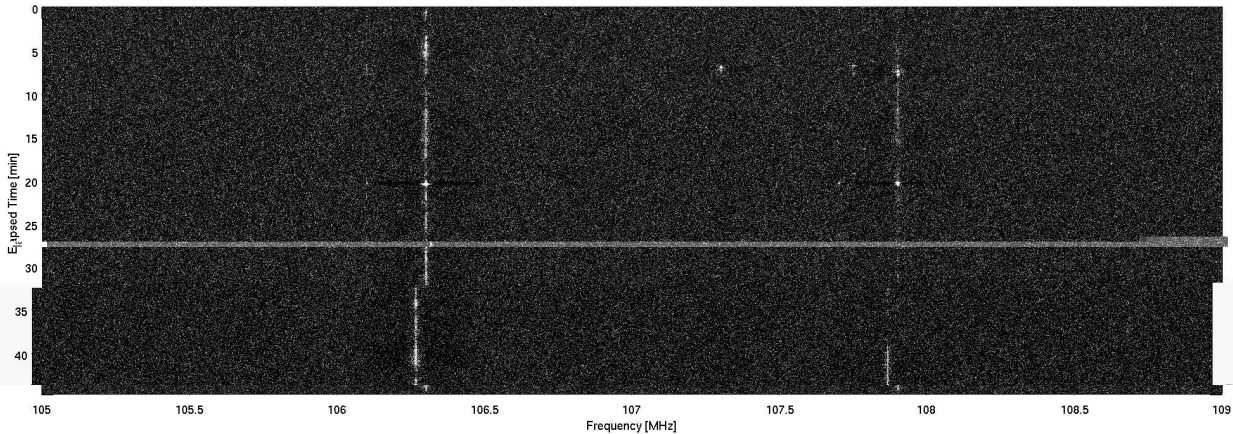} \caption{
\label{f_waterfall} Waterfall plot of RFI in the 4 MHz band spanning
105 to 109 MHz.  The gray scale is such that white corresponds to a
3-$\sigma$ variation in measured power.  Two strong interferers are
visible at 106.3 and 107.9 MHz. Large, short-duration bursts in power
at approximately 7 min and 20 min elapsed time are taken as evidence
of distant transmissions scattering off ionized gas in meteor trails.
}
\end{figure*}

The strong interferers at the frequencies of greatest interest for
EOR experiments (below 200 MHz) tend to be highly time variable.
Figure \ref{f_waterfall} presents a dynamic spectrum or ``waterfall"
plot in order to illustrate the time variability of signals in the
portion of the FM band between 105 and 109 MHz.  Two consistent
interferers with variable amplitudes are observed at 106.3 MHz and
107.9 MHz, and are typically detected at $\sim$15-$\sigma$ in 1 kHz
channels with 1.3 sec integrations. Figure \ref{f_waterfall} also
contains two events at approximately 7 min and 20 min elapsed time
when the power in individual frequency channels, including many
outside the consistent carrier channels, deviates by up to 20 dB, or
well above 100-$\sigma$. Much of the power in the interferers in the
FM band is from short-duration bursts like these. The two interferers
visible in the top panel in Figure \ref{f_rfi_deep} are due entirely
to short-duration events, and the peak at 112.4 MHz in the third
panel is significantly dominated by another two such events, one of
which increases the power in the peak channel by 25 dB.

These events are believed to be due to radiation from distant,
over-the-horizon transmitters scattering off the plasma in meteor
trails toward the array \citep{2001BASI...29..251Y,
1999PASP..111..359M}. Meteoroids enter the Earth's atmosphere at
speeds of order 10 km s$^{-1}$ and ablate significantly around 100 km
altitude.  The ablated atoms interact with molecules in the
atmosphere producing meteor trails, expanding clouds of plasma. Radio
reflections can be produced from radiation scattering off the plasma
immediately surrounding the moving meteoroid or off the trail.
Scattering events from the vicinity of the moving meteoroid are very
brief, $\sim0.1$ sec, while scattering from meteor trails last
generally around one second (although some may persist for minutes).

\begin{figure}
\noindent\includegraphics[width=19pc]{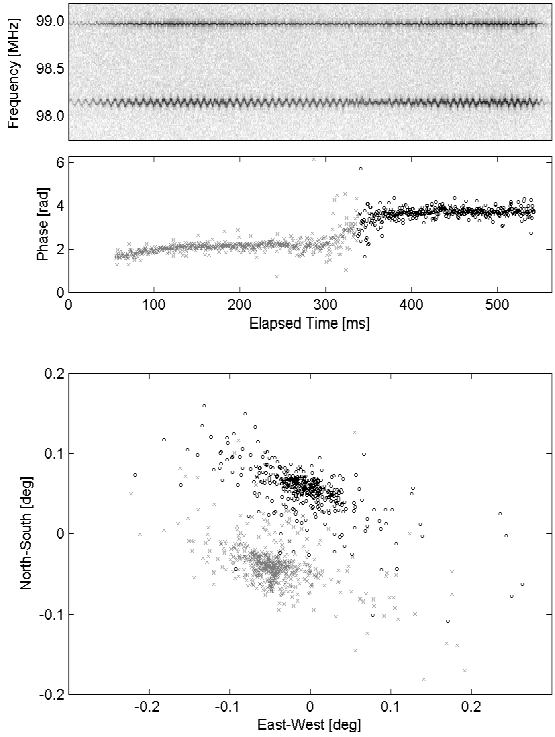} \caption{
\label{f_meteor} Expanded view of signals from two distant
transmitters scattering off ionized gas in a meteor trail.  The top
panel illustrates the high time resolution modulation of the carrier
signals.  The second panel shows the interferometric phase at 98.9
MHz during the length of the reflection event along the baseline
between antenna tiles 1 and 2, and the bottom panel uses the
interferometric phase information from all three baselines to plot
the relative location of the centroid of the reflected emission. The
relative position of the centroid undergoes a clear shift in its
position approximately half way through the event. This observation
was acquired with all three antenna tiles using the north-south
polarization on September 20, 2005 at 21:22:41 UTC.}
\end{figure}

Figure \ref{f_meteor} shows a high time resolution measurement of a
meteor trail event in which carrier waves from two distant
transmitters are briefly visible and seen to modulate with their
audio signals.  All three antenna tiles recorded this event with
sufficient signal to noise that interferometric phases were
available.  Solving for the centroid of the reflected radiation
(Figure \ref{f_meteor}, bottom panel) reveals two distinct
distributions separated by approximately 0.1 deg. At an altitude of
100 km, this angular separation corresponds to an offset of about 175
m. The two distributions are divided in time, as well, with one
coming from the first 300 ms of the event and the other from the
final 200 ms. The transition between the two features lasts
approximately 50 ms. This is consistent with observing the transition
from ``head'' scattering due to plasma immediately surrounding the
meteoroid as it enters the atmosphere to trail scattering due to the
lingering partially ionized cloud that forms in its wake
\citep{2002JGRA.107j.SIA9C}.

The metallic surfaces on aircraft are also capable of reflecting
distant transmissions toward the array. Unlike meteoroids entering
the atmosphere, aircraft are typically moving at speeds of order 0.2
km s$^{-1}$ at altitudes around 30 km; and the rounded surfaces on
aircraft may favorably reflect radiation for up to a few minutes as
the aircraft moves across the sky.

The sporadic nature of the FM signals detected with the prototype
system appears largely to be due to scattering from meteor trails and
reflections from the metallic surfaces on aircraft.  Overall, ten
unique transient events were recorded in the FM band during 2 h of
discontinuous observing. There were seven events lasting less than 3
sec, one event lasting 40 sec, and two events lasting about 65 sec.
The lengths of these events suggest that the seven short-lived
reflections are due to meteor trail reflections and the three longer
events are most likely due to aircraft reflections. The total
temporal occupancy was about $2.5\%$ and the spectral occupancy
during the events increased by as much as a factor of 100 from $<
0.1\%$ to $\sim 1\%$ over a 4 MHz band.

Since the integrated spectrum in the FM band is dominated by meteor
reflections that have very low temporal occupancy, the implication is
that whatever RFI exists in other bands (even well below the
sensitivity threshold achieved in extended integrations by the
prototype system) may be similarly dominated. It may be possible to
eliminate most of the weak RFI in other bands by full-band excision
of time slots when the FM band is observed to contain strong
reflections from over-the-horizon interferers. Using such a scheme,
the MWA--LFD should be able to go much deeper than the -30 dB
achieved with the prototype system, without picking up any
significant interference.

\section{OBSERVATIONAL RESULTS}
\label{s_obs}

\subsection{Fornax A Imaging}

As a final test of the stability of the prototype system and of our
consequent ability to achieve consistent amplitude and phase
calibration over both time and frequency, it is useful to attempt
imaging observations of bright, discrete radio sources. We have done
this for the well-known source For-A (R.A. 03h 22m 40s, Dec.
-36$^{\circ}$ 14' 20"), which has a prominent double morphology of
$\sim$50 arcmin angular extent, sufficient to be resolved by the
prototype array. The total flux density of For-A is $\sim$475 Jy at
150 MHz, rising to $\sim$900 Jy at 80 MHz
\citep{1983A&A...127..361E}.

Data were acquired on September 18, 2005, using 4 MHz bands centered
at 80, 100, 120, 140 and 160 MHz.  The observations cycled between
these and other frequencies every $\sim$7 minutes over a period of
$\sim$10.5 hours, with an on-source integration duty cycle of roughly
3.3\% at each frequency. The data were correlated on-site using the
512-channel mode and the resulting visibility measurements were
stored for later analysis.

The For-A dataset included regular observations of Pic-A, which is
not resolved by the prototype system interferometers.  The
observations provided the gain and phase calibration required in
addition to that needed for the baseline determination (see Section
\ref{s_baseline}) used to correct for geometric delay and Earth
rotation. We also calculated the cross-talk contribution from the
data itself and subtracted it before exporting the data to FITS files
for subsequent analysis and imaging.

Team members experimented with creating images in several imaging
packages. This included the use of self-calibration to correct
phases. All single-frequency images showed clear double structure,
but due to the sparse visibility coverage in the $uv$-plane from only
3 baselines, strong residual sidelobe structures were widespread.

\begin{figure*}
\noindent\includegraphics[width=39pc]{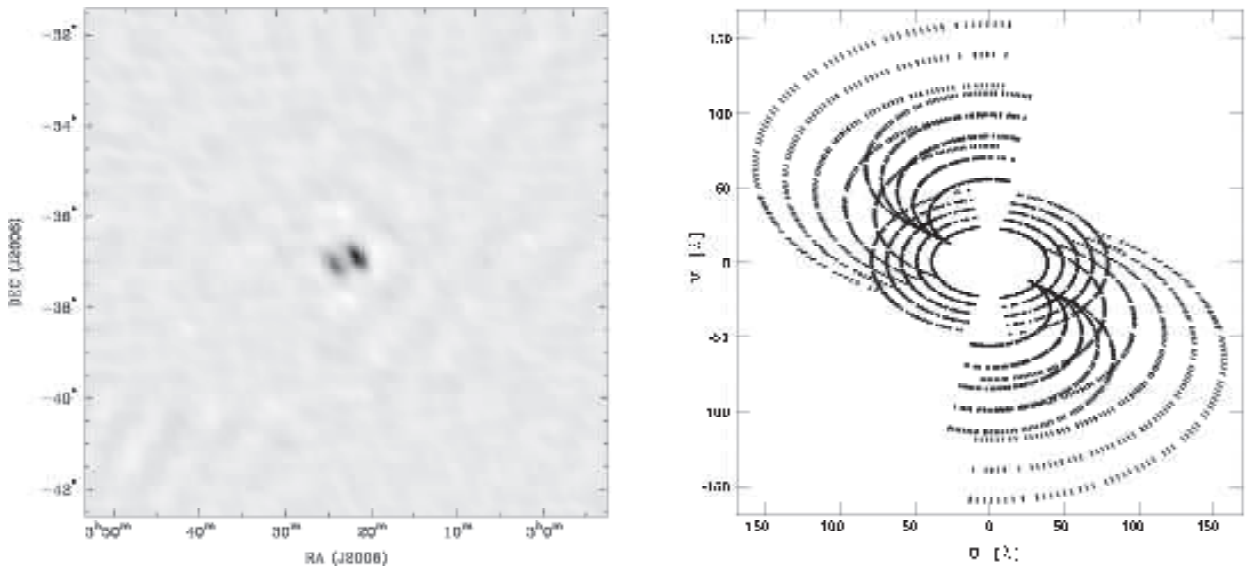} \caption{
\label{f_fornax} Multifrequency synthesis image of For-A (left) and
the distribution of visibility measurements in the $uv$-plane
(right).}
\end{figure*}

In order to improve the image quality, we obtained more complete
$uv$-coverage by implementing ``multifrequency synthesis'' (Sault \&
Conway 1999). The Miriad package turned out to be the most convenient
environment in which to perform this task.  The multifrequency
synthesis algorithm in Miriad generates both an intensity and a
spectral index map of the source, in effect scaling the visibilities
to a compromise flux scale.  The formal dynamic range of the
resulting image in Figure \ref{f_fornax} is $\sim$100:1 (peak/rms),
and the observed structure is in excellent agreement with high
fidelity images of Fornax A at higher frequencies
\citep{1992ApJS...80..137J, 1989ApJ...346L..17F}.

This successful mapping exercise indicates that the calibration of
the prototype antenna tiles and receivers is reasonably well
understood. There are various effects which will limit the fidelity
of the image shown in Figure \ref{f_fornax}, including unmodelled
spectral gradients in the For-A structure, unmodelled time-variable
instrumental response to any linearly polarized emission in For-A,
and the presence of other sources within the antenna tile field of
view. Of these, the biggest source of uncertainty is most certainly
the cumulative effect of other sources entering both the main beam of
the antenna tile and the strong sidelobes, which, in some cases, are
only $\sim 10$ dB below the peak response (see Figure
\ref{f_beammeasurements}). There has been no attempt to perform an
all-sky self-calibration or source subtraction in order to achieve a
more global solution.

\subsection{Solar and Ionosphere Measurements}
\label{s_solar}

During September 12 through 21, 2005, the Sun was observed in an 8
MHz band centered at 100 MHz for several hours per day.  The data
collection mode consisted of an alternating pattern of observing for
64 contiguous seconds and then calculating the auto- and
cross-correlations for all three antenna tiles for a similar amount
of time. This resulted in an approximately 50\% duty cycle. To
maintain a high time resolution in the processed data while
minimizing the downtime for computation, the 512-channel correlator
mode was used to produce 16 kHz-wide frequency channels. The
resulting spectra were saved with 50 ms temporal resolution.

A single, large cluster of sunspots, NOAA active region 0808, was
present on the Sun during the entire prototype field deployment
(March through September, 2005) and was the source of all flares
occurring during the deployment (flare information courtesy of the
NOAA Space Environment Center\footnote{http://sec.noaa.gov/}). During
periods when no sunspots were visible on the solar surface, such as
on September 20, 2005, the measured fluxes at 100 MHz from the Sun
were $\sim 10^4$ Jy.

On September 16, 2005, an M4.4-class flare commenced at 01:41 UTC
from active region 0808.  The active region was located at S11W26 on
the solar disk.  The x-ray flux peaked at 01:49 UTC and decayed to
50\% peak intensity by 01:56 UTC. Observations with the prototype
equipment began at 02:00 UTC and covered the subsequent hour as the
flare continued to decay. In contrast to observations of the quiet
Sun, the measured background power at all frequencies in the 8 MHz
band during this period was elevated by a factor of 40, with typical
values of $\sim 4 \times 10^5$ Jy.

Dozens of short duration radio bursts were observed during this
period.  The bursts had durations of 0.5 sec or less in any one
frequency channel, descended rapidly across the observing band to low
frequencies within seconds, and had peak power levels that could
exceed $5 \times 10^6$ Jy. These are signatures of Type-III solar
radio bursts, which are generated by beams of electrons travelling
along magnetic field lines at speeds above 20,000 km s$^{-1}$
\citep{PlasmaAstrophysics, CairnsKaiser}.

In Type-III solar radio bursts, the motion of the electrons along the
field lines excites fluctuations in the corona at the local plasma
frequency, and these fluctuations are converted into radio emission
at the plasma frequency or at a harmonic. As the electrons escape
along an open field line into interplanetary space, the density of
the local plasma decreases, resulting in a rapid drop in the emitting
frequency. Figure \ref{f_sol_burst} (right panel) shows the measured
power as a function of frequency and time for a clear Type-III burst
observed by the prototype system.

A more detailed study of the properties of the individual radio
bursts will be the subject of a future paper.  In the remainder of
this section we examine the variation of the location of the emission
centroid.

\subsubsection{Emission Centroid}
\label{s_obs_ionosphere}

Since both the bursts and the background solar emission were very
intense during the flare on September 16, 2005, each individual
frequency and time interval had sufficient signal to noise to solve
for the unique location of the centroid of emission using the phases
of the cross-correlated powers. In general, the combination of the
three baselines produced locations repeatedly to within 10 arcsec.

\begin{figure}
\noindent\includegraphics[angle=90,width=19pc]{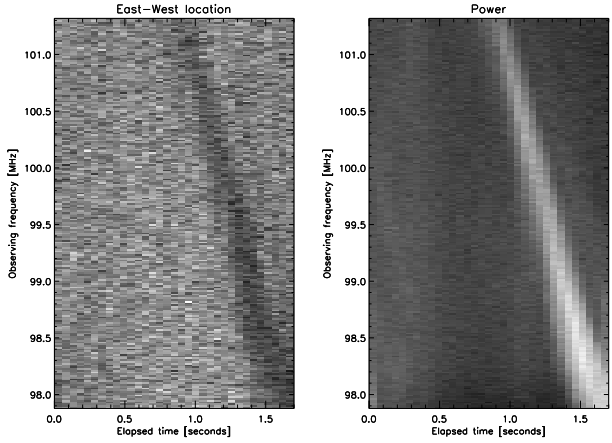} \caption{
\label{f_sol_burst} Grayscale plots of the east-west location (left)
and auto-correlation power (right) as a function of time and
frequency, illustrating an example of a Type-III solar radio burst.
The deflection of the burst is about 20 arcsec, and the burst is
about 20 times more powerful than the background. }
\end{figure}

Two forms of variation were seen.  On a timescale of several hundred
seconds, the location of the centroid for the Sun wandered by about 7
arcmin, much larger than the error in the solutions for the emission
location. Intense radio bursts were associated with sudden
deflections in the source location, as can be seen in Figure
\ref{f_sol_burst} (left panel). We attribute the long-term motion to
small changes in electron column density between each antenna tile
and the Sun due to density gradients in the ionosphere; whereas the
short term change in the emission centroid is believed to be due to
the burst region itself. If the burst region is slightly offset from
the location of the background emission and momentarily dominates the
power along the baselines of the array, the location of the centroid
will move during the bursts.

Figure \ref{f_sol_deflection} is a plot of the location of the
emission centroid as a function of wavelength for one snapshot
interval without radio bursts. The absolute location is unknown due
to uncertainties in the lengths of the cables between the antenna
tiles and other experimental effects, but there is a clear gradient
in any given interval between displacement and wavelength. A simple
model for the angular offset, $\alpha$, that one would expect with
wavelength due to a difference in electron column density in the
ionosphere between the antenna tiles is given by:
\begin{equation}
\label{eqn_tec}sin(\alpha) \sim \frac{\Delta_{TEC}}{\nu^2 d},
\end{equation}
where $\Delta_{TEC}$ is the difference in the total electron content
(TEC), $\nu$ is frequency, and $d$ is the length of the baseline.
Although the observed frequency band was too narrow to clearly see
the $\sim\lambda^2$ dependency expected from Equation \ref{eqn_tec},
the properties of the centroid displacements were consistent overall
with the effects of ionospheric gradients, as opposed to poor
calibration of the baselines, since they exhibited quasi-random
behavior as a function of time. For the observation shown in Figure
\ref{f_sol_deflection}, we identify a best-fit value for the
$\Delta_{TEC}$ of $6.4\times10^{13}$ electrons m$^{-2}$. This is
several thousand times smaller than the typical total electron
content for the daytime ionosphere.

We quantify the ionospheric variation associated with the long-term
(several hundred second) variation in the location of the emission
centroid in Figure \ref{f_sol_histogram}, which shows a histogram of
the relative frequency of occurrence of different values of the
differential electron column density between the antenna tiles. For
the one hour period analyzed, 90\% of the time the difference in TEC
was less than 0.005 TECU, where 1 TECU$\equiv 10^{16}$ electrons
m$^{-2}$ This distribution should represent a relatively extreme case
since the ionosphere was active during this period, as a coronal mass
ejection associated with a flare several days earlier had arrived
recently at Earth and triggered a geomagnetic storm.

\subsubsection{Implications for Ionospheric Calibration}

\begin{figure}
\noindent\includegraphics[angle=90,width=19pc]{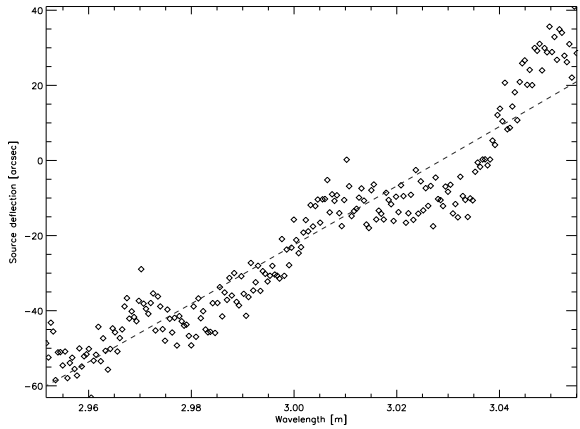} \caption{
\label{f_sol_deflection} Deflection of the centroid of emission as a
function of wavelength for a 50 ms interval without radio bursts. The
dashed line is the best-fit of a model for the deflection expected
due to a difference in electron column density in the ionosphere
between the antenna tiles of $6.4\times10^{13}$ electrons m$^{-2}$. }
\end{figure}

As demonstrated by Figure \ref{f_sol_deflection}, the apparent
positions of astronomical sources on the sky shift due to variations
in electron column density over the array. These positional errors
will contribute to the noise levels in MWA--LFD images and
uncertainties in EOR statistical measurements, thus their calibration
and removal over the wide field of view will be a critical component
of the final image processing path for the full MWA--LFD. With the
antenna tile characteristics measured during the prototype field
deployment, quantitative estimates of the efficacy of the planned
ionospheric calibration can be made.

\begin{figure}
\noindent\includegraphics[angle=90,width=19pc]{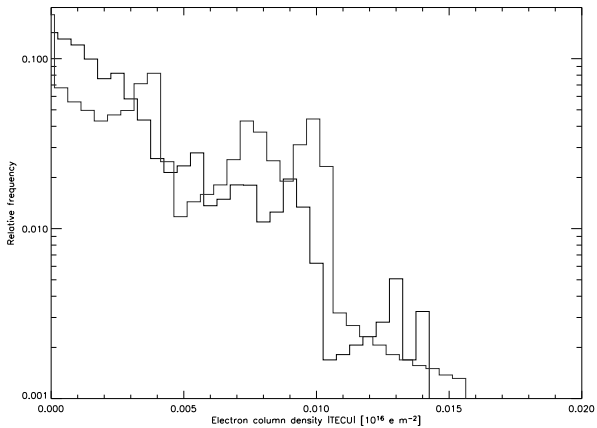} \caption{
\label{f_sol_histogram} Histogram of the relative frequency of
occurrence of differences in electron column density between the
antennas.  The black line is the difference in column density in the
east-west direction, and the gray line is the difference in the
north-south direction. }
\end{figure}

Simulations of the ionospheric calibration process for the full
MWA--LFD have been performed and will be described in detail in a
future paper. The simulated technique is based on algorithm that is
used for Very Large Array data taken at 74 MHz
\citep{2004SPIE.5489..180C, 2004SPIE.5489..354L,
2004ApJS..150..417C}, where the longer baselines and lower frequency
make ionospheric calibration much more challenging than for the
MWA--LFD case.  For a model ionosphere whose electron column density
fluctuations are comparable to those determined in Section
\ref{s_obs_ionosphere} (as shown in Figures \ref{f_sol_deflection}
and \ref{f_sol_histogram}), a 9th-order two-dimensional polynomial
fit to the measured offset in the locations of 160 predetermined
calibrator sources results in RMS residuals in position of 4 to 6
arcsec (2 to 3\% of the 200 MHz array beam) over an 8$^\circ$ by
8$^\circ$ field of view. The residual noise level was $F \cong 10$
$\mu$Jy, equivalent to the thermal noise level that would be reached
by the MWA--LFD in $\sim 1000$ h of integration. Although the field
of view of the actual array will be greater than 20$^\circ$ by
20$^\circ$ for frequencies below 200 MHz---and the model fit would be
according less accurate given the same number of calibrator sources
and order of polynomial---with refinement it is projected that
ionospheric calibration will not limit MWA--LFD image fidelity or
significantly interfere with EOR statistical measurements.

\section{CONCLUSION}
\label{s_concl}

The design of the MWA--LFD is a departure from traditional approaches
to radio astronomy instrumentation. The wide field of view and large
bandwidth coverage of the instrument, combined with the phased-array
design of the antenna tiles and low-cost digital receiver system,
create exciting new opportunities for scientific exploration in a
part of the electromagnetic spectrum that has been largely neglected
for decades. But for the array to fully succeed in achieving its
science objectives requires an understanding of the operational
properties of the instrument at unprecedented levels.

The field testing campaign reported on here was conducted to begin
addressing this requirement at an early stage in the development of
the array. The results of the effort demonstrated the fundamental
performance of the MWA--LFD prototype antenna tiles and receiver
system in the Western Australia environment.  The experiments further
indicated that the RFI environment of the site is excellent and
should not pose a significant hurdle for the planned observations,
including those targeting the epoch of reionization.

Analysis of the data obtained with the prototype system confirmed the
presence of several Type-III solar bursts during the field campaign
and probed their spectral, temporal, and spatial structure;
constrained the properties of the local ionosphere during a period of
high solar activity; and detected several bright giant pulses from
the Crab Nebula pulsar. These results are all relevant to active
fields of study and provide enticing precursors of what will be
achievable with the full array.

\acknowledgments

Many people and institutions have contributed to the MWA--LFD project
and the prototype field deployment effort. Contributing institutions
include Massachusetts Institute of Technology, MIT Kavli Institute
for Astrophysics and Space Research, and MIT Haystack Observatory,
University of Melbourne, Australian National University, Curtin
University, Australian National Telescope Facility, University of
Western Australia, Harvard-Smithsonian Center for Astrophysics,
Mileura Cattle Company, the government of Western Australia, and the
Australian Research Council.

Contributing individuals include: Duncan Campbell-Wilson, Priscilla
Clayton, Rob Gates, Lincoln Greenhill, Karen Haines, Rich Jackson,
Kelly Kranz-Little, Tony Martin-Jones, Alan Rogers, Michelle Storey,
Lian Walsh, and Patrick Walsh.

This work was supported by Australian Research Council, the MIT
School of Science, and the MIT Haystack Observatory. Low frequency
astronomy, heliospheric science, and radio array technology
development at MIT are supported by the National Science Foundation.

%ARC: grant DP0345001%

We would also like to thank the Australian Partnership for Advanced
Computing National Facility for use of the Mass Data Storage System
at the ANU Supercomputer Facility.

\bibliography{apjmnemonic,../../../references}
\bibliographystyle{apj}

\begin{deluxetable}{lcc}
    \tablecaption{Antenna Tile Locations}
    \tablewidth{0pt}
    \tablehead{
        \colhead{} &
        \colhead{Equatorial ($x, y, z$) [m]} &
        \colhead{Local ($x_{topo}, y_{topo}, z_{topo}$) [m]}}
    \startdata
        Tile 1 & (0, 0, 0) & (0, 0, 0) \\
        Tile 2 & (-1.08, 145.64, -2.56) & (0.175, 145.64, -2.77) \\
        Tile 3 & (-76.43, 278.45, -149.28) & (-1.993, 278.45, -167.70)
    \enddata
    \tablecomments{ \label{t_tiles}
    Antenna tile locations determined by observations of unresolved
    continuum sources at 120 and 275 MHz.  The left set of locations is
    in conventional coordinates for interferometer baselines, described
    in Section \ref{s_baseline}, and the lower set is in a topocentric
    coordinate system, where $x_{topo}$ is aligned with up, $y_{topo}$ is
    aligned with east, and $z_{topo}$ completes a right-handed system by
    pointing into the north. }
\end{deluxetable}

\begin{deluxetable}{lcc}
    \tablecaption{Baseline Determination Comparison}
    \tablewidth{0pt}
    \tablehead{
        \colhead{} &
        \colhead{Celestial [m]} &
        \colhead{Tape [m]}}
    \startdata
    Tile 1 - Tile 2 & 145.67 & 145.85 \\
    Tile 1 - Tile 3 & 325.06 & 325.2 \\
    Tile 2 - Tile 3 & 211.76 & 211.55
    \enddata
    \tablecomments{ \label{t_comparison} Comparison of the baselines determined using
    astronomical observations and those measured by tape.}
\end{deluxetable}

\begin{deluxetable}{lc}
    \tablecaption{Site Location}
    \tablewidth{0pt}
    \tablehead{
        %\colhead{} &
        %\colhead{}
        }
    \startdata
    Latitude & 26$^{\circ}$ 25' 52" S \\
    Longitude & 117$^{\circ}$ 12' 24" E \\
    Elevation & 427 m
    \enddata
    \tablecomments{ \label{t_location}Latitude and longitude of the prototype field deployment site (determined by the Global Positioning System (GPS)).}
\end{deluxetable}

\begin{deluxetable}{lcccccccccccc}
    \tablecaption{Mileura Weather}
    \tablewidth{0pt}
    \tablehead{
        \colhead{} &
        \colhead{Jan} &
        \colhead{Feb} &
        \colhead{Mar} &
        \colhead{Apr} &
        \colhead{May} &
        \colhead{Jun} &
        \colhead{Jul} &
        \colhead{Aug} &
        \colhead{Sep} &
        \colhead{Oct} &
        \colhead{Nov} &
        \colhead{Dec}
        }
    \startdata
    Mean High Temp [C] & 37.8 & 36.8 & 34.0 & 29.0 & 23.2 & 19.1 & 18.4 & 20.4 & 24.6 & 28.3 & 32.8 & 36.2\\
    Mean Low Temp [C] & 22.8 & 22.4 & 20.1 & 15.7 & 11.1 & 8.2 & 6.9 & 7.8 & 10.1 & 13.1 & 17.2 & 20.7 \\
    Mean Rainfall [mm] & 25.4 & 29.8 & 22.9 & 19.6 & 25.5 & 29.1 & 25.0 & 17.8 & 6.9 & 6.7 & 8.4 & 14.0
    \enddata
    \tablecomments{\label{t_climate}Climate averages at Cue, Western Australia.  From the Australian Government Bureau of Meteorology.}

%% (http://www.bom.gov.au/climate/averages/tables/cw_007017.shtml)

\end{deluxetable}

\end{document}